\documentclass{llncs} %

\usepackage{float}
\usepackage{graphicx}

\usepackage[labelfont=bf]{caption}
\usepackage{tabularx}
\usepackage{color}
\usepackage{amsmath} 
\usepackage{mathtools} 
\usepackage{amsfonts} 
\usepackage{amssymb}
\usepackage[normalem]{ulem}

\usepackage{array}
\usepackage{booktabs}
\usepackage{rotating}
\usepackage{multirow}
\usepackage{multicol}
\usepackage{lscape}
\usepackage{subfig}
\usepackage{comment}

\usepackage{wrapfig}
\usepackage{rotating}
\usepackage{epstopdf}

\usepackage{lineno}
\usepackage{soul}
\usepackage{color}
\usepackage{xcolor}
\usepackage{xargs}

\usepackage[export]{adjustbox}

\usepackage{algorithm,algorithmicx,algpseudocode}

\usepackage{color}

\definecolor{brilliantrose}{rgb}{1.0, 0.33, 0.64}

\usepackage[colorlinks=true,linkcolor=blue,citecolor=blue]{hyperref}

\usepackage[colorinlistoftodos,prependcaption,textsize=tiny]{todonotes}
\newcommandx{\ISLEM}[2][1=]{\todo[linecolor=brilliantrose,backgroundcolor=brilliantrose!25,bordercolor=brilliantrose,#1]{#2}}

\usepackage{tikz,xcolor,hyperref}

\definecolor{lime}{HTML}{A6CE39}
\DeclareRobustCommand{\orcidicon}{
	\begin{tikzpicture}
	\draw[lime, fill=lime] (0,0) 
	circle [radius=0.16] 
	node[white] {{\fontfamily{qag}\selectfont \tiny ID}};
	\draw[white, fill=white] (-0.0625,0.095) 
	circle [radius=0.007];
	\end{tikzpicture}
	\hspace{-2mm}
}

\foreach \x in {A, ..., Z}{\expandafter\xdef\csname orcid\x\endcsname{\noexpand\href{https://orcid.org/\csname orcidauthor\x\endcsname}
			{\noexpand\orcidicon}}
}
			
\usepackage{color}

\definecolor{darkgreen}{rgb}{0.53, 0.66, 0.42}

\begin{document}

\title{Foreseeing Brain Graph Evolution Over Time Using Deep Adversarial Network Normalizer}

\titlerunning{Short Title}  

\author{Zeynep G\"{u}rler\index{G\"{u}rler, Zeynep}\inst{1}, Ahmed Nebli\orcidB{}\inst{1,2} \and Islem Rekik\orcidA{} \index{Rekik, Islem}\inst{1}\thanks{ {corresponding author: irekik@itu.edu.tr, \url{http://basira-lab.com}. } This work is accepted for publication in the PRedictive Intelligence in MEdicine (PRIME) workshop Springer proceedings in conjunction with MICCAI 2020 }}

\institute{$^{1}$ BASIRA Lab, Faculty of Computer and Informatics, Istanbul Technical University, Istanbul, Turkey \\ $^{2}$ National School for Computer Science (ENSI), Mannouba, Tunisia }

\authorrunning{Z. G\"{u}rler et al.}

\maketitle              

\begin{abstract}

Foreseeing the brain evolution as a complex highly interconnected system, widely modeled as a graph, is crucial for mapping dynamic interactions between different anatomical regions of interest (ROIs) in health and disease. Interestingly, brain \emph{graph} evolution models remain almost absent in the literature. Here we design an adversarial brain network normalizer for representing each brain network as a transformation of a fixed centered population-driven connectional template. Such graph normalization with respect to a fixed reference  paves the way for reliably identifying the most similar training samples (i.e., brain graphs) to the testing sample at baseline timepoint. The testing evolution trajectory will be then spanned by the selected training graphs and their corresponding evolution trajectories. We base our prediction framework on geometric deep learning which naturally operates on graphs and nicely preserves their topological properties. Specifically, we propose the first \emph{graph-based} Generative Adversarial Network (gGAN) that not only learns how to normalize brain graphs with respect to a fixed connectional brain template (CBT) (i.e., a brain template that selectively captures the most common features across a brain population) but also learns a high-order representation of the brain graphs also called embeddings. We use these embeddings to compute the similarity between training and testing subjects which allows us to pick the closest training subjects at baseline timepoint to predict the evolution of the testing brain graph over time. A series of benchmarks against several comparison methods showed that our proposed method achieved the lowest brain disease evolution prediction error using a single baseline timepoint. Our gGAN code is available at \url{http://github.com/basiralab/gGAN}.

\keywords{adversarial network normalizer $\cdot$ brain graph evolution prediction $\cdot$ connectional brain template $\cdot$  graph generative adversarial network $\cdot$ sample selection}

\end{abstract}

\section{Introduction}

Early disease diagnosis using machine learning has become the new essence of modern-day medicine. Studies have shown that predicting the evolution of brain diseases can dramatically change the course of treatment and thus maximizing the chance of improving patient outcome \cite{Querbes:2009}. For instance, \cite{leifer2003,grober1995} found that neurodegenerative diseases such as dementia are no longer reversible if diagnosed at a late stage. In this context, several research papers have attempted to combine neuroimaging with the predictive robustness of deep learning frameworks. As such, in one study, \cite{payan2015} used 3D convolutional neural networks to predict the onset of Alzheimer's disease (AD). However, these studies relied on samples that were taken at late disease stages which cannot be useful for prescribing \emph{personalized} treatments for patients. To address this limitation, we are interested in solving a more challenging problem which is  predicting the evolution of a brain disease over time given only an initial timepoint.

Previous studies have developed shape-based and image-based prediction frameworks using morphological features derived from brain MRI scans to foresee the brain evolution trajectory \cite{Gafuroglu:2018,Rekik:2017c}. For instance, \cite{Rekik:2017c} used a representative shape selection method to predict longitudinal development of cortical surfaces and white matter fibers assuming that similar shapes at baseline timepoint will have similar developmental trajectories. Such an assumption has been also adopted in a landmark study \cite{Gafuroglu:2018}, demonstrating the reliability of exploring similarities between baseline training and testing samples for predicting the evolution of brain MR image trajectory in patients diagnosed with mild cognitive impairment.  Although these works proposed successful predictive frameworks for image-based brain evolution trajectory prediction and classification, these were solely restricted to investigating the brain as a surface or a 3D image. This undeniably overlooks the integral and rich representation of the brain as a \emph{graph}, where the pairwise interconnectedness between pairs of anatomical regions of interest (ROIs) is investigated. To overcome this limitation, \cite{Ezzine:2019} proposed a Learning-guided Infinite Network Atlas selection (LINAs) framework, the first study that designed a learning-based atlas-to-atlas similarity estimation to predict brain graph evolution trajectory over time solely from a single observation. Despite its promising prediction accuracy, in the sample selection step, LINAs first vectorized each brain graph by storing the connectivity weights in the graph adjacency matrix in a feature vector. This fails to preserve the brain graph topology since the vectorization step regards the graph as a Euclidean object. A second limitation of these works is that such sample connectomic representation via vectorization might include irrelevant and redundant features that could mislead the training sample selection step.

To address these limitations, we tap into the nascent field of geometric deep learning aiming to learn representations of non-Euclidean objects such as graphs while preserving their geometry. Drawing inspiration from previous brain evolution predictive frameworks \cite{Gafuroglu:2018,Rekik:2017c,Ezzine:2019}, we also assume the preservation of local sample (i.e., brain graph) neighborhood  across different timepoints. As such, by \emph{learning} the similarities between samples at baseline timepoint, one can identify the most similar training brain graphs to a testing brain graph. By integrating the evolution trajectories of the selected training samples, one can then predict the evolution trajectory of the testing brain graph. To this aim, we model each training and testing graph as a deformation or a transformation of a fixed reference, namely a connectional brain template (CBT). Such hypothesis is inspired from the classical deformable theory template widely adopted in Euclidean image-based registration frameworks \cite{Allassonniere:2015,Trouve:1995}, 
where each sample, in this case an image, is represented as a diffeomorphic transformation of a fixed template. 

Specifically, we design the first \emph{graph-based} generative adversarial network (gGAN) \cite{goodfellow2014} that learns how to \emph{normalize} a brain graph with respect to a fixed connectional brain template (CBT). A CBT can be viewed as a center of a population of brain graphs as proposed in \cite{Dhifallah:2020}, selectively capturing the most common features across population brain graphs. Our gGAN is composed of a graph normalizer network that learns a high-order representation of each brain graph as it gets transformed into a fixed CBT, thereby producing a CBT-based \emph{normalized} brain graph. Our gGAN normalizer is also coupled with an adversarial CBT-guided discriminator which learns how to differentiate between a normalized brain network and the reference CBT. We use our trained normalizer's weights to embed both subjects' brain graphs and the fixed CBT. Next, we compute the residual between each training normalized sample embedding and the target testing normalized sample embedding to eventually identify the most similar training samples for the target prediction task. Below, we articulate the main contributions of our work at different levels:

\begin{enumerate}

\item We propose to model each brain graph observation as a transformed version  of a population graph template. Hence, each brain graph can be normalized with respect to the fixed template, thereby producing a more individualized brain graph capturing its unique and individual connectivity patterns.  

\item  We propose the first gGAN that learns how to normalize a set of graphs with respect to a fixed biological connectional template (i.e., a CBT).

\item  Our prediction framework of brain network evolution trajectory is a generic framework. Hence, it can be used to foresee both healthy and atypical evolution of brain connectivity from a single timepoint.

\end{enumerate}

\begin{sidewaysfigure}
\includegraphics[width=19.5cm]{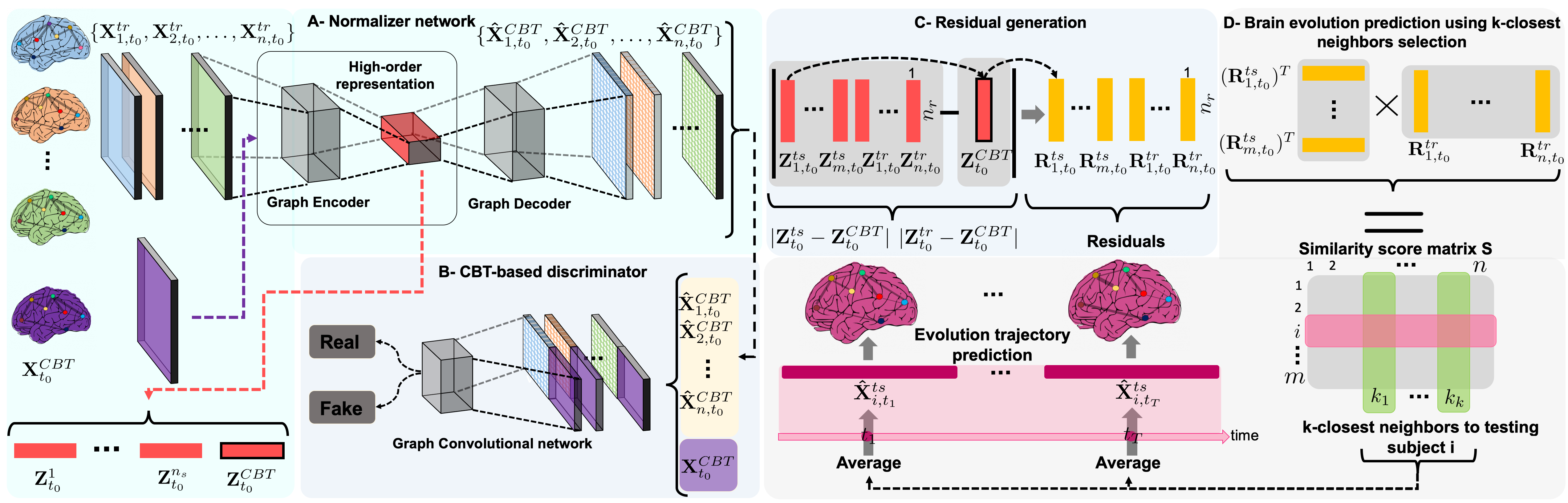} 
\caption{ \emph{Proposed gGAN based sample selection strategy for predicting brain network evolution from baseline timepoint $t_0$.} \textbf{(A)} \emph{Normalizer network}. We develop a gGAN that learns to normalize brain graphs with respect to a fixed connectional brain template (CBT). To do so, we design a three-layer graph convolutional neural network normalizer acting as an encoder and a decoder mimicking a U-net architecture. Our normalizer takes a set of $n$ training subjects $\mathbf{X}_{t_0}^{tr}$ at timepoint $t_0$ and outputs a set of $\mathbf{\hat{X}}_{t_0}^{CBT}$ that aim to share the same distribution as the population CBT. We use the learned weights from the normalizer's first two layers to embed a set of $n_s$ subjects as well as the CBT. \textbf{(B)} \emph{CBT-based discriminator}. We design a two-layer graph convolutional neural network that differentiates between the real CBT $\mathbf{X}_{t_0}^{CBT}$ and the \emph{normalized} brain graph $\mathbf{\hat{X}}_{t_0}^{CBT}$. \textbf{(C)} \emph{Residual Generation.} The CBT embedding $\mathbf{Z}_{t_0}^{CBT}$ is subtracted from the testing embeddings $\mathbf{Z}_{t_0}^{ts}$ and training embeddings $\mathbf{Z}_{t_0}^{tr}$, respectively, to generate testing residuals $\mathbf{R}_{t_0}^{ts}$ and training residuals $\mathbf{R}_{t_0}^{tr}$. \textbf{(D)} \emph{Brain graph evolution prediction using $k$-closest neighbors selection.} A similarity score matrix is generated by multiplying the training residuals $\mathbf{R}_{t_0}^{tr}$ by the transpose of the testing residuals $(\mathbf{R}_{t_0}^{ts})^{T}$ to compute the dot product (similarity) between training subjects $\mathbf{X}^{tr}$ and testing subjects $\mathbf{X}^{ts}$. To predict the brain graph evolution of subject $i$ over timepoints $\{t_1,\dots,t_T\}$, we select the top $k$ training subjects with the highest similarity scores to the baseline testing subject to predict its evolution trajectory $\{\mathbf{\hat{X}}_{i,t_1}^{ts},\dots,\mathbf{\hat{X}}_{i,t_T}^{ts}\}$ by taking the average of these neighboring graphs at $\{t_1,\dots,t_T\}$.}
\label{fig:1}
\end{sidewaysfigure}

\section{Proposed Method}

In this section, we introduce the key steps of our gGAN-based sample selection strategy for predicting brain graph evolution trajectory over time from a single timepoint. \textbf{Table}~\ref{tab:1} displays the mathematical notations that we use throughout this paper. We denote the matrices as boldface capital letters, e.g., $\mathbf{X}$, and scalars as lowercase letters, e.g., $n$. The transpose operator is denoted as $\mathbf{X}^{T}$. In the following sections, we will detail each of the four key steps of our prediction framework as shown in \textbf{Fig.}~\ref{fig:1}.

\textbf{Connectional brain template (CBT) generation.} A CBT is a brain graph template that holds the most shared, representative, and centered brain connectivities across a population of brain graphs. It was first introduced by \cite{Dhifallah:2020} as an efficient framework to identify discriminative features that help spot out  disordered brain connections by comparing healthy and disordered CBTs. Here, we first set out to define the fixed CBT to integrate into our gGAN architecture using an \emph{independent} brain graph dataset composed of $n_c$ subjects.

Let $\mathbf{V}_{(i,j)}^{s}$ denote the pairwise connectivity between ROIs $i$ and $j$ of a subject $s$; $1 \leq i,j \leq n_r$. For each pair of ROIs $i$ and $j$, we define a high-order graph $\mathbf{H}_{(i,j)}$ $\in\mathbb{R}^{n_c \times n_c}$ that holds the pairwise distances across all subjects for each pair of ROIs $(i,j)$ as follows:

\begin{gather}
    \mathbf{H}_{(i,j)}(s, s^{\prime}) = \sqrt{(\mathbf{V}_{(i,j)}^{s} - \mathbf{V}_{(i,j)}^{s^{\prime}})^{2}}; \ \forall \ 1 \ \leq s, s^{\prime} \leq n_c 
\end{gather}

Next, we construct a distance vector $\mathbf{M}_{(i,j)}(s)$ for each subject $s$ that computes the cumulative distance between subject $s$ and other subjects in the independent set for connectivity $(i,j)$. $\mathbf{M}_{(i,j)}(s)$ can be regarded as the topological strength of node $s$ in the high-order graph $\mathbf{H}_{(i,j)}$.
 
\begin{gather}
    \mathbf{M}_{(i,j)}(s) = \sum_{s^{\prime}=1}^{n_c} \mathbf{H}_{(i,j)}(s, s^{\prime}) = \sum_{s^{\prime}=1}^{n_c} \sqrt{(\mathbf{V}_{(i,j)}^{s} - \mathbf{V}_{(i,j)}^{s^{\prime}})^{2}}; \ \forall 1 \ \leq s, s^{\prime} \leq n_c 
\end{gather}

Finally, for each brain connectivity $(i,j)$, we select the connectivity weight of the subject achieving the minimum cumulative distance to all other subjects with the assumption that the closest subject's connectivity to all other subjects is indeed the most representative and centered one. Therefore, we define the independent population CBT as follows:

\begin{gather}
    \mathbf{X}^{CBT}_{(i,j)} = \mathbf{V}_{(i,j)}^{k}; \ where \ k = \min\limits_{1 \leq s \leq n_c} \mathbf{M}_{(i,j)}(s)
\end{gather}

\textbf{Overview of CBT-guided prediction of brain graph evolution framework from baseline.} GANs are deep learning frameworks composed of two neural networks: a generator $G$ and a discriminator $D$ \cite{goodfellow2014}. The generator is an encoder and decoder neural network aiming to learn how to generate fake data output that mimics the original data distribution while the discriminator learns how to differentiate between the ground truth data and the fake data produced by the generator. These two networks compete against each other in an adversarial way so that with enough training cycles, the generator learns how to generate more \emph{real-looking} fake samples and the discriminator learns to better discriminate between the real and fake samples. Since this framework has proven its efficiency in translating input data into the desired output domain  (e.g., translating T1-MRI to T2-MRI \cite{yang2020}), we propose to modify the generator's task from fake sample production to a \emph{normalization-based} mapping learning from an input space nesting brain graphs to a \emph{fixed} template (i.e., a CBT ); and hence, we call it the normalizer network $N$. To the best of our knowledge, our proposed framework is the first gGAN composed of a graph normalizer network, mapping to a fixed output, and a discriminator.

\begin{center}
\begin{table}
	\begin{scriptsize}
\caption{\label{tab:1} Major mathematical notations}
\resizebox{\textwidth}{!}{\begin{tabular}{c c}
 \hline
 \textbf{Mathematical notation}   & \hspace{1 cm} \textbf{Definition}  \\
 \hline
 $n_s$    &  number of subjects for training and testing our model  \\
 $n$      & total number of training subjects  \\
 $m$      & total number of testing subjects  \\ 
 $n_r$    & total number of regions of interest in brain \\
 $n_c$      & total number of independent subjects for CBT generation \\
 $\mathbf{V}^{s}$   & brain connectivity matrix of subject $s$ \\
 $\mathbf{H}_{(i,j)}$   & high-order graph $\in\mathbb{R}^{n_c \times n_c}$ defined for a pair of ROIs $i$ and $j$ \\
 $\mathbf{M}_{(i,j)}(s)$ & node strength of subject $s$ in the high-order graph $\mathbf{H}_{(i,j)}$ \\
 $\mathbf{X}^{CBT}$ & connectional brain template connectivity matrix  \\
 $\mathbf{X}_{t_0}^{tr} = \{\mathbf{X}_{1,t_0}^{tr}, \dots,\mathbf{X}_{n,t_0}^{tr}\}$ &  training brain graph connectivity matrices $\in\mathbb{R}^{n\times n_r \times n_r}$ at $t_0$ \\
 $\mathbf{X}_{t_0}^{ts} = \{\mathbf{X}_{1,t_0}^{ts}, \dots,\mathbf{X}_{m,t_0}^{ts}\}$  &  testing brain graph connectivity matrices $\in\mathbb{R}^{m\times n_r \times n_r}$ at $t_0$ \\
 $\mathbf{\hat{X}}_{t_0}^{CBT} = \{\mathbf{\hat{X}}_{1,t_0}^{CBT}, \dots,\mathbf{\hat{X}}_{n,t_0}^{CBT}\}$  &  CBT-normalized training connectivity matrices $\in\mathbb{R}^{n\times n_r \times n_r}$ at $t_0$ \\
 $N$ & GAN normalizer  \\
 $D$ & GAN CBT-guided discriminator  \\
 $\mathcal{L}_{full}$ & full loss function \\
 $\mathcal{L}_{adv}$ & adversarial loss function \\
 $\mathcal{L}_{L_1}$ & $l_1$ loss function \\
 $\lambda$ & coefficient of $l_1$ loss \\
 $V$  & a set of $n_r$ nodes \\
 $E$  & a set of $m_r$ directed or undirected edges \\
 $l$  & index of layer \\
 $Y^{l}$ & transformation matrix $\in\mathbb{R}^{n_r\times d_l }$ \\
 $L$  & transformation matrix $\in\mathbb{R}^{m_r\times d_m}$ \\
 $\mathcal{N}(i)$ & the neighborhood containing all the adjacent nodes of node $i$ \\
 $Y^{l}(i)$ & filtered signal of node $i$ $\in\mathbb{R}^{d_l}$ \\
 $F^{l}_{ji}$ & filter generating network \\
 $\omega^{l}$ & weight parameter \\
 $b^{l}$ & bias parameter \\
 $\mathbf{Z}_{t_0}^{tr} = \{\mathbf{Z}_{1,t_0}^{tr}, \dots,\mathbf{Z}_{n,t_0}^{tr}\}$  & training brain graph embeddings $\in\mathbb{R}^{n\times n_r}$ at $t_0$ \\
 $\mathbf{Z}_{t_0}^{ts} = \{\mathbf{Z}_{1,t_0}^{ts}, \dots,\mathbf{Z}_{m,t_0}^{ts}\}$  & testing brain graph  embeddings $\in\mathbb{R}^{m\times n_r}$ at $t_0$ \\
 $\mathbf{Z}_{t_0}^{CBT}$  &  CBT embedding $\in\mathbb{R}^{n_r}$ at $t_0$ \\
 $\mathbf{S}$              & similarity score matrix $\in\mathbb{R}^{m \times n}$ \\
$\mathbf{S}_{i}$          & similarity score vector of testing subject $i$ $\in\mathbb{R}^{n}$ \\
 $\mathbf{S}_{i,j}$        & similarity score between testing subject $i$ and  training subject $j$ \\
 $\mathbf{R}_{t_0}^{tr} = \{\mathbf{R}_{1,t_0}^{tr}, \dots,\mathbf{R}_{n,t_0}^{tr}\}$  & residuals of embedded training subjects at $t_0$ $\in\mathbb{R}^{n \times n_r}$ \\
 $\mathbf{R}_{t_0}^{ts} = \{\mathbf{R}_{1,t_0}^{ts}, \dots,\mathbf{R}_{m,t_0}^{ts}\}$  & residuals of embedded testing subjects at $t_0$ $\in\mathbb{R}^{m \times n_r}$ \\
 $\mathbf{\hat{X}}_{i,t_1}^{ts}$    & predicted test subject $i$ at $t_1$ $\in\mathbb{R}^{n_r \times n_r}$ \\
 $\mathbf{\hat{X}}_{i,t_T}^{ts}$    & predicted test subject $i$ at $t_T$ $\in\mathbb{R}^{n_r \times n_r}$\\
 \hline
\end{tabular}}
	\end{scriptsize}
\end{table}
\end{center}

First, we start by training our gGAN to learn how to normalize brain graphs of a set of $n$ training subjects $\mathbf{X}_{t_0}^{tr}$ at timepoint $t_0$ with respect to a fixed CBT. This will enable us to map each subject's brain graph into a fixed CBT, thereby producing each CBT-normalized brain graph $\mathbf{\hat{X}}_{t_0}^{CBT}$. We use the learned weights from our normalizer's encoding block to embed training subject $\mathbf{X}_{t_0}^{tr}$ and testing subject $\mathbf{X}_{t_0}^{ts}$. We also feed the CBT as an input to the normalizer network to produce a self-normalized embedding $\mathbf{Z}_{t_0}^{CBT}$. Next, for each training subject, we calculate its residual embedding with respect to the CBT by taking the absolute difference between the CBT embedding $\mathbf{Z}_{t_0}^{CBT}$ and the subject normalized embedding $\mathbf{Z}_{t_0}^{tr}$. We also produce similar residual embeddings for the testing subjects.  We then use these residual embeddings to define a similarity score matrix computing the dot product between a pair of training and testing residual embeddings  $\mathbf{R}_{t_0}^{tr}$ and $\mathbf{R}_{t_0}^{ts}$ (\textbf{Fig.}~\ref{fig:1}--D). Note that this boils down to computing the cosine similarity between two vectors with unitary norms. Finally, for each testing subject, we select the top $k$ training subjects with the highest similarity scores, and predict the evolution trajectory by simply averaging their corresponding training trajectories. 

Our gGAN aims to optimize the following loss function: 
\begin{gather}
 arg min_{N} max_{D} \mathcal{L}_{adv} = \mathbb{E}_{x \sim p_{(CBT)}} [log D(x) ] + \mathbb{E}_{\hat{x}  \sim p_{(\mathbf{X^{tr}})}  }[log( 1- D(N(\hat{x}))) ] 
\end{gather}

To improve the quality of the CBT-normalized brain graph, we propose to preserve each subject's embedding scheme by adding an $L_1$ loss term that minimizes the distance between each normalized subject $\hat{\mathbf{X}}^{tr}_{t_{0}}$ and its related ground-truth brain graph $\mathbf{X}^{tr}_{t_{0}}$. Therefore our full loss function is expressed as follows:

\begin{gather}
\mathcal{L}_{full} = \mathcal{L}_{adv} + \lambda \mathcal{L}_{L1}(N) 
\end{gather}

\textbf{The normalizer network}. As shown in \textbf{Fig.}~\ref{fig:1}--A, our proposed normalizer network is composed of three-layer graph convolutional neural network (GCN) inspired by the dynamic edge convolution operation introduced in \cite{SimonovskyK17} and mimicking a U-net architecture \cite{Olaf:2015} with skip connections that enhance brain graph normalization and thus improve the quality of our normalized graph embeddings \cite{mao2016}. The normalizer takes a set of $\mathbf{X}^{tr}_{t_{0}}$ training subjects as input and outputs a set of $\mathbf{\hat{X}}^{CBT}_{t_{0}}$ which share the same distribution as the fixed CBT.  Hence, our normalizer's encoder not only learns a deep non-linear mapping between any subject's brain graph and the fixed reference graph (i.e., CBT) but also a high-order embedding of the input with regard to the CBT.  

Our normalizer contains three graph convolutional neural network layers regularized using batch normalization \cite{ioffe2015} and dropout \cite{xiao2016} to the output of each layer. These two operations undeniably help simplify and optimize the network training. For instance, batch normalization was proven to accelerate network training through a rapid convergence of the loss function while dropout was proven to eliminate the risk of overfitting.

\textbf{CBT-guided discriminator}. We display the architecture of the discriminator in \textbf{Fig.}~\ref{fig:1}--B. The discriminator is also a graph neural network inspired by \cite{SimonovskyK17}. Our proposed discriminator is a two-layer graph neural network that takes as input a concatenation of the normalizer's output $\mathbf{\hat{X}}^{CBT}_{ t_{0}}$ and the CBT. The discriminator outputs a value between $0$ and $1$ characterizing \emph{the realness} of the normalizer's output. To improve our discriminator's ability to differentiate between the fixed CBT and CBT-normalized samples, we design our gGAN's loss function so that it maximizes the discriminator's output value for the CBT and minimize it for each $\mathbf{\hat{X}}^{CBT}_{ t_{0}}$.

\textbf{Dynamic graph-based edge convolution}. Each of the graph convolutional layers of our gGAN architecture uses a dynamic graph-based edge convolution operation proposed by \cite{SimonovskyK17}. In particular, let $G=(V, E)$ be a directed or undirected graph where $V$ is a set of $n_{r}$ ROIs and $E  \subseteq V \times V $ is a set of $m_{r}$ edges. Let $l$ be the layer index in the neural network.  We define $Y^{l}: V \rightarrow \mathbb{R}^{d_{l}}$ and $L: E \rightarrow \mathbb{R}^{d_m}$ which can be respectively considered as two transformation matrices (i.e., functions) where $Y^{l} \in \mathbb{R}^{n_{r} \times d_{l}}$ and $L \in \mathbb{R}^{m_{r} \times d_m}$. $d_m$ and $d_{l}$ are dimensionality indexes. We define by $\mathcal{N}(i)= \left\{ j; (j, i) \in E \right\} \cup \left\{i \right\}  $ of a node $i$ the neighborhood containing all the adjacent ROIs. 

The goal of each layer in both the normalizer and the discriminator is to output the graph convolution result which can be considered as a filtered signal $Y^{l}(i) \in \mathbb{R}^{d_{l}}$ at node $i$. $Y^{l}$ is expressed as follows:

\begin{gather*}
Y^{l} (i) = \frac{1}{\mathcal{N}(i)} \sum_{j \in \mathcal{N}(i)} \Theta^{l}_{ji} Y^{l-1} (j) + b^{l}, 
\end{gather*}

where $\Theta^{l}_{ji} = F^{l} (L(j, i); \omega^{l})$. We note that $F^{l}: \mathbb{R}^{d_m} \rightarrow \mathbb{R}^{d_{l} \times d_{l}-1}$ is the filter generating network, $\omega^{l} $ and $b^{l}$ are model parameters that are updated only during training. \par

\textbf{Embedding the training, testing subjects and the CBT.} We recall that our gGAN's main purpose is to (i) learn how to normalize brain graphs with respect to a fixed CBT and (ii) learn a CBT-normalized embedding. As shown in \textbf{Fig}~\ref{fig:1}--A, once we train the normalizer network using our training set, we produce the embeddings of the training subjects, testing subjects, and the CBT (i.e., self-embedding). We define $\mathbf{Z}_{t_0}^{tr}$ and $ \mathbf{Z}_{t_0}^{ts}$ as the results of our embedding operation of training and testing data, respectively. Given that our normalizer encodes brain graphs and extracts their high-order representative features  in a low-dimensional space with respect to the CBT, we assume that such embeddings might be better representations of the brain graphs as they capture individual traits that distinguish them from the population `average'.

\textbf{Residual computation and sample similarity estimation.} As shown in \textbf{Fig}~\ref{fig:1}--C, we obtain the residuals between the embedding of each brain graph and the CBT embedding by calculating their  absolute differences. Next, we use these residuals to define the similarity score matrix $\mathbf{S}$ $\in\mathbb{R}^{n \times m},$ where each element $\mathbf{S}_{i,j}$ expresses the pairwise similarity between a row-wise testing subject $\mathbf{X}_{i}^{ts}$ and a column-wise training subject $\mathbf{X}_{j}^{tr}$. Specifically, to obtain the similarity matrix, we calculate the dot product of the matrix composed of the vertically stacked transposed residual embeddings of testing subjects and the matrix composed of the vertically stacked residuals of training subjects. As stated in \cite{ding1999}, the dot product of two normalized matrices provides the similarity between them.  As a result, the greater the value of the element of the similarity matrix is, the most similar the related subjects are. We note the training and testing residuals as $\mathbf{R}_{t_0}^{tr}$ and $ \mathbf{R}_{t_0}^{ts}$, respectively, and we define them as follows:

\begin{gather}
    \mathbf{R}_{ t_0}^{tr} = |\mathbf{Z}_{t_0}^{tr} - \mathbf{Z}_{t_0}^{CBT}| \\   
    \mathbf{R}_{ t_0}^{ts} = |\mathbf{Z}_{t_0}^{ts} - \mathbf{Z}_{t_0}^{CBT}| 
\end{gather}

\textbf{Brain graph evolution prediction using top $\textbf{k}$-closest neighbor selection.}  
Assuming that the top $k$-closest neighbors of the testing subjects will remain neighbors at the following timepoints $t \in \{t_1, \dots, t_T \}$ \cite{Rekik:2017c,Gafuroglu:2018}, we predict the brain graph evolution by selecting its most similar $k$ training subjects at baseline. Next, we predict the testing subject's brain evolution by averaging its corresponding training subjects' graphs (i.e., neighbors) at follow-up timepoints. We select the top $k$ subjects for each testing subject using their highest corresponding elements in the similarity score matrix. To predict the evolution of a baseline testing brain graph $i$, we sort its derived row $\mathbf{S}_{i}$ vector in the similarity score matrix $\mathbf{S}$ and select the top $k$-samples with the highest similarity scores. Given a baseline testing brain graph $\mathbf{X}^{ts}_{i,t_0}$, we foresee the evolution of its connectivity $\mathbf{\hat{X}}_{i,t}^{ts}$ at later timepoints $t \in \{t_1, \dots, t_T \}$  by averaging the $k$ selected training brain graphs at each timepoint $t$.

\section{Results and Discussion}

\textbf{Evaluation dataset.} We used $114$ subjects from the OASIS-2\footnote{\url{https://www.oasis-brains.org/}} longitudinal dataset \cite{Marcus:2010}. This set consists of a longitudinal collection of 150 subjects aged 60 to 96. Each subject was scanned on two or more visits, separated by at least one year. For each subject, we construct a cortical morphological network derived from cortical thickness measure using structural T1-w MRI as proposed in \cite{Mahjoub:2018}. Each cortical hemisphere is parcellated into $35$ ROIs using Desikan-Killiany cortical atlas. We built our gGAN with PyTorch Geometric library \cite{Pytorchgeometric:2019} and trained it using 3-fold cross-validation applied on $n = 91$ training subjects. We randomly selected $n_c = 23$ subjects from the OASIS-2 dataset \cite{Marcus:2010} to generate a CBT using the netNorm \cite{Dhifallah:2020}.

\textbf{Parameter setting.} We varied the number of selected neighboring samples $k$ from $\{2, \dots, 10\}$ for the target prediction task. In \textbf{Table} 2, we report prediction mean absolute error averaged across $k$. We set the normalizer's loss hyperparameter  to $100$ which is $\times100$ the adversarial loss. Also, we chose ADAM \cite{Adam:2014} as our default optimizer and set the learning rate at $0.001$ for the normalizer and $0.01$ for the discriminator. We set the exponential decay rate for the first moment estimates (e.i., beta 1) to $0.5$, and the exponential decay rate for the second-moment estimates (e.i., beta 2) to $0.999$ for the ADAM optimizer. Finally, we trained our gGAN for $700$ epochs using NVIDIA Tesla V100 GPU.

\textbf{Comparison methods and evaluation.} We benchmarked our framework against three comparison methods for neighboring sample selection (SS) using: (i) the original graph features (OF) which is a baseline method that computes the dot product similarities between \emph{vectorized} connectivity matrices of testing and training graphs as in \cite{Ezzine:2019}.  (ii) CBT-based residuals (SS-CR), which is a variation of SS-OF where we first produce residuals by computing the absolute difference between the \emph{vectorized} brain graphs and the \emph{vectorized} CBT, then compute the dot product between the produced residuals of training and testing subjects. Note that in these two variants, we are not producing any embeddings of the brain graphs. (iii) CBT-normalized embeddings (SS-CE), which is a variant of our method that discards the residual generation step (\textbf{Fig.}~\ref{fig:1}--C) and predicts the brain graph evolution by computing the dot product between the embeddings of the training graphs and the testing graphs by gGAN. 

All benchmarks were performed by calculating the mean absolute error (MAE) between the ground-truth and predicted brain graphs of the testing subjects at $t_1$ and $t_2$ timepoints and varying the number of selected training samples $k$ in the range of $\{2, \dots, 10\}$ for a better evaluation. \textbf{Table} 2 shows the results of MAE-based prediction accuracy for $t_1$ and $t_2$ timepoints.

\begin{table}
\begin{center}

\caption{ Prediction accuracy using mean absolute error (MAE) of our proposed method and comparison methods at $t_1$ and $t_2$ timepoints. We report the MAE averaged across $k \in \{2, \dots, 10\}$.}

\resizebox{\textwidth}{!}{\begin{tabular}{c | c | c | c | c}
\hline
 & \multicolumn{2}{c|}{$t_1$} & \multicolumn{2}{c}{$t_2$} \\
 \hline
 \textbf{Method}   
 &  \begin{tabular}{c c}
    \textbf{Mean MAE}   \\
     $\pm$ std
 \end{tabular} 
 & \begin{tabular}{c c}
     \textbf{Best}   \\
     \textbf{MAE}
 \end{tabular}
 &   \begin{tabular}{c c}
     \textbf{Mean MAE}   \\
     $\pm$ std
 \end{tabular} 
 &  \begin{tabular}{c c}
     \textbf{Best}   \\
     \textbf{MAE}
 \end{tabular} \\
 \hline
  SS-OF &  \hspace{0.2 cm}$0.04469 \pm 0.00247$  \hspace{0.2 cm} &  \hspace{0.2 cm}$0.04194$ \hspace{0.2 cm} &  \hspace{0.2 cm} $0.05368 \pm 0.00449$ \hspace{0.2 cm} & \hspace{0.2 cm} $0.04825$ \hspace{0.2 cm} \\
 SS-CR & $0.04417 \pm 0.002026$ & $0.04225$ & $0.05045 \pm 0.000942$ & $0.04939$ \\
 SS-CE & $0.04255 \pm 0.001835$ & $\mathbf{0.04064}$ & $0.04948 \pm 0.002480$ & $0.04707$ \\
 Ours & $\mathbf{0.04237 \pm 0.001679}$ & $0.04075$ & $\mathbf{0.04882 \pm 0.002517}$ & $\mathbf{0.04624}$ \\

 \hline
\end{tabular}}
\end{center}
\end{table}

Our proposed brain graph framework integrating both CBT-based normalization and CBT-based residual computation steps outperformed baseline methods at both timepoints. Our method also achieved the best MAE in foreseeing the brain graph evolution at $t_2$. However, the best MAE for prediction at $t_1$ was achieved by SS-CE, which uses the gGAN normalizer network and discards the residual computation with respect to the CBT. This might be due to the fact that subjects are more likely to be more divergent from the center at $t_2$ than $t_1$. Overall, our sample selection using CBT-guided embedded residuals achieved the best performance in foreseeing brain graph evolution trajectory and showed that normalizing brain graphs with respect to a fixed graph template such as a CBT is indeed a successful strategy outperforming methods using the original brain graphs. 

\textbf{Limitations and future work.} Although our graph prediction framework achieved the lowest average MAE against benchmarking methods in predicting brain graph evolution trajectory from a single observation, it has a few limitations. So far, the proposed method only handles uni-modal brain graphs with a single edge type. In our future work, we aim to generalize our gGAN normalizer to handle brain multigraphs.  In a multigraph representation of the brain wiring, the interaction between two anatomical regions of interest, namely the multigraph nodes, is encoded in a set of edges of multiple types. Each edge type is defined using a particular measure for modeling the relationship between brain ROIs such as functional connectivity derived from resting state functional MRI or morphological similarity derived from structural T1-weighted MRI. Furthermore, our framework can only operate on undirected and positive brain graphs. Extending our framework to handle directed and signed networks would constitute a big leap in generalizing our approach to different biological and connectomic datasets.

\section{Conclusion}

In this paper, we proposed a novel brain graph evolution trajectory prediction framework based on a gGAN architecture comprising a normalizer network with respect to a fixed connectional brain template (CBT) to first learn a topology-preserving (using graph convolutional layers) brain graph representation. We formalized the prediction task as a sample selection task based on the idea of using the residual distance of each sample from a fixed population center (CBT) to capture the unique and individual connectivity patterns of each subject in the population.  Our results showed that our brain graph prediction framework from baseline can remarkably boost the prediction accuracy compared to the baseline methods. Our framework is generic and can be used in predicting both typical and disordered brain evolution trajectories. Hence, in our future work we will evaluate our framework on large-scale connectomic datasets with various brain disorders such as brain dementia. We will investigate the potential of \emph{predicted} evolution trajectories in boosting neurological disordered diagnosis.

\section{Supplementary material}

We provide three supplementary items for reproducible and open science:

\begin{enumerate}
	\item A 6-mn YouTube video explaining how our prediction framework works on BASIRA YouTube channel at \url{https://youtu.be/5vpQIFzf2Go}.
	\item gGAN code in Python on GitHub at \url{https://github.com/basiralab/gGAN}. 
	\item A GitHub video code demo on BASIRA YouTube channel at \url{https://youtu.be/2zKle7GzrIM}. 
\end{enumerate}

\section{Acknowledgement}

I. Rekik is supported by the European Union's Horizon 2020 research and innovation programme under the Marie Sklodowska-Curie Individual Fellowship grant agreement No 101003403 (\url{http://basira-lab.com/normnets/}).

\bibliography{Biblio3}
\bibliographystyle{splncs}
\end{document}